\documentclass[10pt,twocolumn]{IEEEtran}
\usepackage{amssymb, algorithm, algpseudocode}
\usepackage{amsmath}
\usepackage{siunitx}
\usepackage{setspace} 
\usepackage{widetext}
\usepackage{algorithmicx}
\IEEEoverridecommandlockouts
\usepackage[active]{srcltx} 
\usepackage{graphicx}
\usepackage{balance}
\usepackage{cuted}
\usepackage{mathtools}
\usepackage{cite}
\usepackage{amssymb}
\usepackage{amsfonts}
\usepackage{amsthm}
\usepackage{hyperref}
\usepackage{braket}
\usepackage{balance}
\usepackage{mathtools}
\usepackage{cite}
\usepackage{graphicx}
\usepackage{textcomp}

\newtheorem{definition}{\textbf{Definition}}
\usepackage{xcolor}
\usepackage{ulem}
\makeatletter
\renewcommand*\env@matrix[1][*\c@MaxMatrixCols c]{%
  \hskip -\arraycolsep
  \let\@ifnextchar\new@ifnextchar
  \array{#1}}
\makeatother

\newcommand{\ketbra}[2]{\vert #1 \rangle  \langle #2 \vert}

\def\BibTeX{{\rm B\kern-.05em{\sc i\kern-.025em b}\kern-.08em
    T\kern-.1667em\lower.7ex\hbox{E}\kern-.125emX}}
\usepackage[T1]{fontenc}
\usepackage{hyperref}
\usepackage{algpseudocode}
\usepackage{subcaption}
\usepackage{ulem}
\usepackage{braket}
\usepackage{tikz}
\usetikzlibrary{quantikz}
\usepackage{adjustbox}
\usepackage{easyReview}
\usepackage{rotating}
\usepackage{placeins}
\usepackage{dblfloatfix}

\begin{document}

\newcommand{\meterCom}{\scalebox{.5}{\begin{tikzcd} \meter{} \end{tikzcd}}}
\newcommand{\cwCom}{\scalebox{.5}{\begin{tikzcd} \pgfmatrixnextcell \cw \end{tikzcd}}}

\newcommand{\eqdef}{\stackrel{\triangle}{=}}


\newtheorem{theo}{Theorem}
\newtheorem{theor}{Theorem}
\newtheorem{cor}{Corollary}
\newtheorem{lem}{Lemma}
\newtheorem{prop}{Proposition}

\renewcommand{\qed}{$\blacksquare$}

\renewcommand{\algorithmiccomment}[1]{// #1}

\title{Space-Based Quantum Internet: Entanglement Distribution in  Time-Varying LEO Constellations
}

\author{Seid Koudia*, Junaid ur Rehman and Symeon Chatzinoatas ~\IEEEmembership{Fellow,~IEEE}
	\thanks{The authors are with the Interdisciplinary Centre for Security, Reliability, and Trust (SnT), Luxembourg, L-1855 Luxembourg. 
		(emails: \{seid.koudia,  junaid.urrehman, symeon.chatzinotas\}@uni.lu)}
}


\maketitle

\begin{abstract}
This paper addresses the complexities of entanglement distribution in LEO satellite networks, particularly those arising from their dynamic topology. Traditional static and dynamic entanglement distribution methods often result in high entanglement drop rates and reduced end-to-end throughput. We introduce a novel framework that leverages the dynamic nature of LEO satellite networks to enhance entanglement distribution efficiency. Employing a space-time graph model to represent the network's temporal evolution, we propose an entanglement distribution strategy based on path utility, incorporating pointing errors, non-ideal link transmittance for intersatellite links, and atmospheric effects for downlinks. Our approach demonstrates superior performance in reducing entanglement drop rates and improving throughput compared to conventional methods. This study advances the field of quantum communication in satellite networks, offering resilient and efficient entanglement distribution strategies that support practical applications such as distributed computing, quantum multipartite cryptography, and distributed quantum sensing. The findings underscore the potential of integrating dynamic satellite networks with quantum technologies to create a reliable and secure quantum internet.
\end{abstract}


\section{introduction}
Satellite communications stand as a cornerstone of modern connectivity, extending the reach of global communication networks beyond terrestrial limitations. In the realm of quantum technology, the potential for satellite-based systems to revolutionize secure communication and enable the quantum internet is increasingly evident \cite{Wehner2018QuantumIA}.

Quantum networks, leveraging the principles of quantum mechanics, offer unparalleled security in communication through protocols such as Quantum Key Distribution (QKD) \cite{kisseleff2023trusted, koudia2024spatial} and entanglement-based schemes \cite{koudia2023quantum, koudia2023deterministic}. These networks promise a future where information transfer is not only secure but also fundamentally quantum, enabling applications like quantum teleportation and distributed quantum computing \cite{kozlowski2019towards}.

However, the implementation of quantum networks faces unique challenges \cite{koudia2024physical}, particularly in extending their reach over vast distances and ensuring robustness in dynamic environments. Satellite networks emerge as a crucial component in overcoming these challenges \cite{Sym-nongeostrationary-survey,sym-nonterrestrial-survey}. By harnessing the capabilities of satellites, quantum communication can transcend geographical constraints and establish global-scale networks with unprecedented security \cite{junaid-survey}.

One key characteristic of satellite networks, especially those operating in Low Earth Orbit (LEO) systems, is their time-varying topology. The dynamic nature of satellite constellations introduces fluctuations in connectivity, including path loss variations, and link configurations over time \cite{Sym-Megaconstellations,junaid2023estimating}. Understanding and effectively managing these variations are essential for the reliable distribution of quantum entanglement, a cornerstone of quantum communication protocols \cite{Jun-quantum-opt-for-leo}.

The aim of this paper is to address the challenges associated with entanglement distribution in LEO satellite constellations, particularly those stemming from the time-varying topology of such networks. Traditional static or dynamic entanglement distribution methods often suffer from high entanglement drop rates, leading to a significant decrease in end-to-end entanglement throughput \cite{optimal}. This paper introduces a novel framework that leverages the dynamic nature of LEO satellite networks to enhance performance rather than hinder it. By adopting a space-time graph model to represent the network's temporal evolution, the paper proposes an entanglement distribution strategy based on path utility. This approach assesses the utility of point-to-point entanglement links within the network, factoring in pointing errors, non-ideal link transmittance for intersatellite connections, and atmospheric effects for downlinks. This innovative method proves to be more efficient and reliable than traditional dynamic distribution approaches, particularly in terms of reducing entanglement drop rates and improving entanglement throughput. Moreover, depending on the coherence times of the quantum memories utilized within the network, this new scheme can achieve high levels of end-to-end entanglement fidelity. By transforming the inherent challenges of a time-varying topology into advantages, this paper presents a robust solution for entanglement distribution in satellite-based quantum networks.

The paper is organized as follows: Section~\ref{sec:2} presents the state of the art in entanglement distribution within satellite-based quantum networks, highlighting the novelty of our approach. Section~\ref{sec:3} details the space-time model employed for our entanglement distribution method, along with other physical models underlying the satellite-based quantum network. In Section~\ref{sec:4}, we discuss the outage probability and utility functions for both the downlinks and intersatellite links. Section~\ref{sec:5} introduces our Space-Time Based Entanglement Distribution (STBD) scheme. In Section~\ref{sec:6}, we analyze the scheme's performance, comparing various network metrics against dynamic routing entanglement distribution conducted over single snapshots. Finally, we conclude the paper with a summary and future perspectives.

\section{State of the art of Entanglement distribution in Space-based Quantum Networks}
\label{sec:2}
\subsection{General Entanglement Distribution}

Entanglement distribution is a critical component of quantum communication networks. It involves the process of creating and sharing entangled quantum states between distant nodes in a network. Over the years, significant advancements have been made in the methods and technologies used for entanglement distribution. These advancements include the development of quantum repeaters, which are used to extend the distance over which entanglement can be distributed by mitigating the effects of decoherence and loss. Quantum repeaters use a combination of entanglement swapping and purification to maintain high fidelity over long distances.

Quantum entanglement routing has been the focus of numerous studies in recent years. In \cite{van2013path}, a routing scheme was proposed with the end-to-end entanglement rate as the key metric for optimality. In \cite{chen}, a greedy algorithm was proposed for finding optimal paths optimizing end-to-end entanglement rates on quantum memory limited paths. A multi-source multi-destination framework has been the focus of \cite{Li}, with optimal paths considering end-to-end entanglement fidelity as a figure of merit.  The work presented in \cite{Moe} introduces a remote entanglement distribution scheme for linear repeater chains, determining an optimal route that maximizes end-to-end entanglement and the optimal sequence for entanglement swaps. Distributed entanglement routing algorithms that consider latency are the focus of \cite{dahlberg}. The challenge of optimizing end-to-end entanglement in scenarios with multiple source-destination pairs is examined in detail as a multicommodity flow problem in \cite{wehner-multico}. Furthermore, \cite{Gyongyosi_2019} develops a decentralized adaptive routing scheme that accounts for the imperfections of quantum memories. Additionally, \cite{Pirandola_multi} explores a multipath routing approach for multiple end-to-end entanglement, providing a comprehensive analysis of this strategy. These approaches consider static network topologies. 

Static routing in satellite networks involves predetermined paths for entanglement distribution \cite{liorni2021quantum,TNO, thales} \footnote{\color{black}Static entanglement distribution is referring to scenarios where the entanglement distribution path is fixed between the two end points and is not adapted to the satellites movement, or inter-satellite links are not considered at all}. These paths are established based on the initial configuration and remain fixed regardless of the changing network topology due to the movement of satellites. Static routing is relatively simple to implement and requires minimal computational resources. However, it is not adaptive to the dynamic nature of satellite networks, which can lead to suboptimal performance and increased entanglement drop rates. In scenarios where the satellite constellation is not optimized for specific ground station pairs, static routing may fail to provide consistent and high-quality entanglement distribution.

\subsection{Entanglement Distribution in Space-Based Networks}

The deployment of quantum communication networks in space, particularly using Low Earth Orbit (LEO) satellites, has opened new frontiers for entanglement distribution. Space-based quantum communication leverages the advantages of the reduced atmospheric absorption and scattering effects in the vacuum of space, allowing for long-distance entanglement distribution without the limitations of terrestrial optical fibers.

Significant milestones have been achieved with the launch of quantum communication satellites, such as the Chinese Micius satellite, which demonstrated successful entanglement distribution over 1200 kilometers \cite{micius}. These experiments have paved the way for global quantum communication networks, linking distant ground stations via satellite-based entanglement distribution.

In \cite{Khatri}, the authors propose a comprehensive model for such a network, emphasizing continuous, on-demand entanglement distribution to ground stations. Their technique optimizes satellite configurations to ensure continuous coverage while balancing the total number of satellites and entanglement distribution rates. Similarly, \cite{Towsley} focuses on the optimal distribution of bipartite entanglement between pairs of ground stations, characterizing the best satellite-to-ground station transmission scheduling policy to maximize the aggregate entanglement distribution rate under various resource constraints, akin to dynamic routing. In \cite{optimal}, the authors delve into optimizing the allocation of quantum entanglement resources to enhance network efficiency and reliability, focusing on the network's response to multiple entanglement requests between ground stations. Additionally, \cite{dynamic} presents a model that accounts for satellite movement and inter-satellite links over time, using a logical graph to represent the network's temporal connections. They formulate an optimization algorithm for entanglement distribution within this dynamic framework, ensuring robust and adaptive strategies. These studies collectively advance the understanding and implementation of entanglement distribution in satellite-based quantum networks, addressing continuous distribution, optimal scheduling, dynamic routing, and time-dependent models to contribute to the development of a reliable and efficient quantum internet.

\section{The Model}
\label{sec:3}
\subsection{Modeling the Time-Varying Topology}

\begin{figure*}[htbp]
\centering
    \begin{minipage}{1\textwidth}
    \includegraphics[width=1\textwidth]{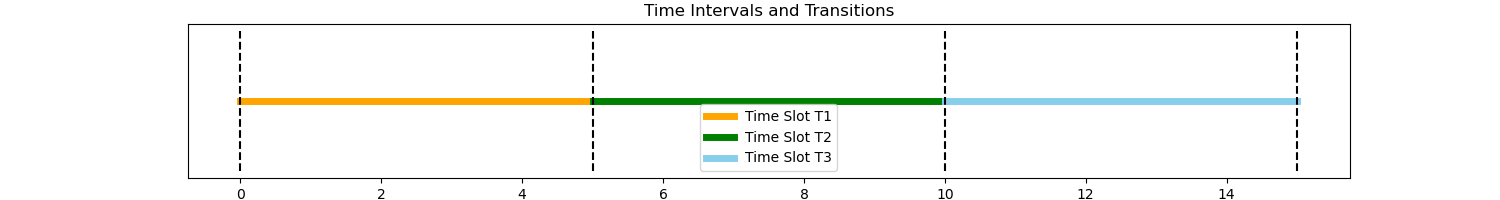}
    \subcaption{The management time of the satellite communication $T$ is splitted into time slots referred to as snapshots}
    \label{fig:time slots}
    \end{minipage}
    \vspace{10pt}
    \begin{minipage}{0.8\textwidth}
    \includegraphics[width=1\textwidth]{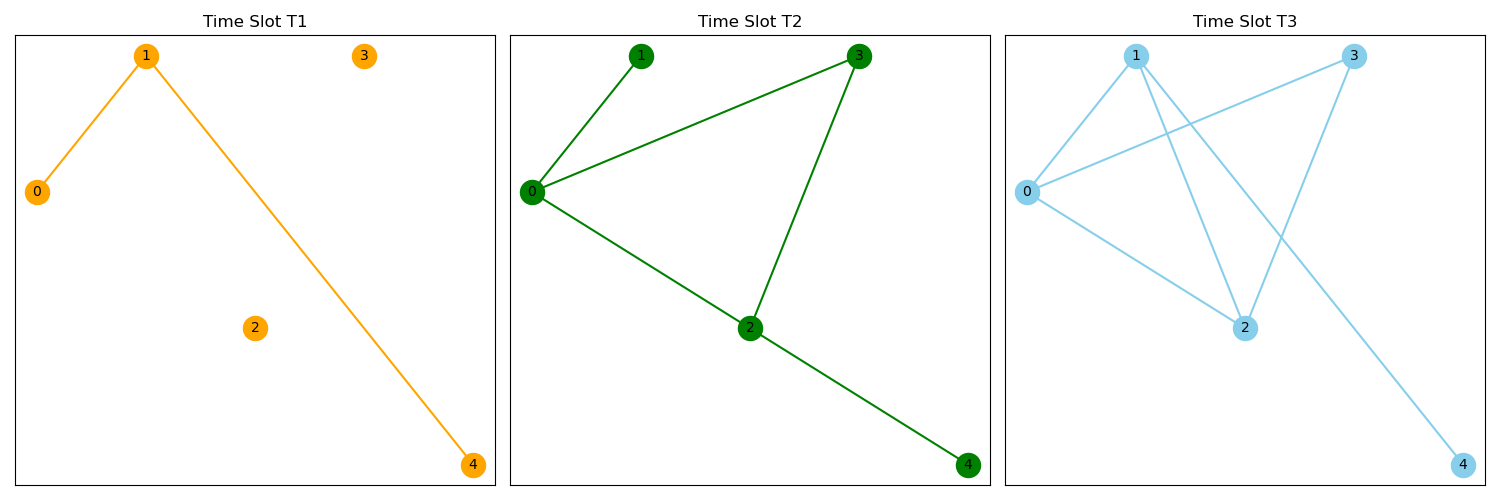}
    \subcaption{The changing network connectivity in each snapshot}
    \label{fig:snapshot graphs}
    \end{minipage}
    \vspace{10pt}
    \begin{minipage}{0.8\textwidth}
    \centering
    \includegraphics[width=0.6\textwidth]{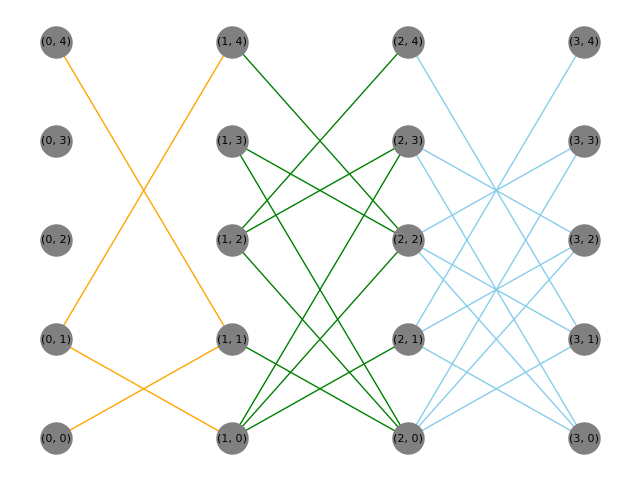}
    \subcaption{The space-time graph describing the network topology change during the management time}
    \label{fig:snapshot graphs}
    \end{minipage}
\caption{A full description of the space-time graph model used in the paper. (a) The time management of the satellite network is splitted into time intervals referred to as snapshots. Whithin each snapshot the satellite topology is assumed not changing as highlighted in (b), giving some coarse graining of the network evolution. (c) The space-time graph describing the underlying time-varying topology of the physical satellite network, reflecting the connectivities in each snapshot. The transition points between snapshot 1 and 2 are 1 and 4, and the transition points between snapshot 2 and 3 are 4, 3, 2, 1 and 0. The weights of the each link in the virtual network are elaborated in Eq.~\ref{weights}.}
\label{illustrating figure}
\end{figure*}

In quantum satellite communication systems, the efficiency of entanglement distribution relies heavily on the underlying network topology. To address the inherent dynamism of quantum satellite networks, we propose a time-varying topology model \cite{zhang2019efficient}. This model encapsulates the dynamic evolution of satellite connectivity through a sequence of discrete graphs representing different network states over time. Leveraging the predictability inherent in satellite movements, we introduce a weighted time-space evolution graph. This graph effectively illustrates the progressive transformation of the time-varying topology, providing insights into its temporal dynamics and facilitating efficient routing strategies. Such modeling is feasible within software-defined LEO satellite networks, allowing an SDN controller to orchestrate entanglement distribution decisions by leveraging real-time topology data.

\begin{definition}
    \textit{The space-time graph model:} Let $V_N=\{v_1, v_2, \cdots, v_N\}$ be the set of nodes in the satellite network, where $N$ is the number of nodes, representing different satellites or ground stations. Assume that time is divided into discrete time slots $\{T_1, T_2, \cdots, T_M\}$, where $M$ is the number of time slots. Each time slot corresponds to a snapshot of the network topology. The snapshot is associated with a time interval, not a specific time instance.
\end{definition}

\begin{definition}
    \textit{Time intervals and transitions:} Let the topology management time $T$ be divided into discrete time slots $T=\{T_1, T_2, \cdots, T_M\}$. Each time slot $T_m$ is a time set consisting of time points $t$. If $t'$ is the initial point of a given time slot, we call it a transition point. Let each time slot be defined as a time interval $T_m=[t_{m-1}, t_m)$, with the set of all transition points being $\{t_0, \cdots, t_M\}$.
\end{definition}

\begin{definition}
    \textit{Snapshots:} The predictable network topology during a time slot $T_m=[t_{m-1}, t_m)$ can be described by a directed graph $G^m=(V^m, E^m)$, where $V_N^m$ is the set of nodes during the time slot and a link $v_i^m v_j^m \in E^m$ represents that nodes $v_i^m$ and $v_j^m$ can exchange entanglement in the time slot $T_m=[t_{m-1}, t_m)$. The evolution of the network can be described by the sequence of snapshots $(S^1, \cdots, S^m, \cdots, S^M)$. The transition point $t_m$ represents the time point when the network topology evolves from snapshot $S^m$ to $S^{m+1}$.
\end{definition}

Based on the space-time model introduced, the time dimension can be integrated into the description of the virtual topology of satellite networks. Benefiting from the periodicity and predictability of the line of sight (LoS) connectivity of satellite networks, the topology within time interval  $[t_0, t_M]$ can be divided into a series of snapshots. The full description of the model is illustrated in Fig.~\ref{illustrating figure}

\subsection{Other Physical Considerations of the Network}

On each established link of the time-varying topology, we assume an EPR pair of the form 
\begin{equation}
    \ket{\Phi^+}=\frac{1}{\sqrt{2}}(\ket{00}+\ket{11})
\end{equation}
is distributed. The noise model considered for propagation is the depolarizing channel, which maps the distributed EPR pair on link $x$ into the noisy Werner state:
\begin{equation}
    \rho_{W_x}=W_x \ketbra{\Phi^+}{\Phi^+}+(1-W_x)\frac{\mathrm{I}_4}{4}
\end{equation}

The fidelity of this noisy state is given in terms of the Werner parameter $W_x$ as:
\begin{equation}
    F=\frac{1+3W_x}{4}
\end{equation}
\textcolor{black}{with $W\in [0,1]$}. 
We consider the first generation of quantum repeaters, which only distribute EPR pairs and perform local Bell State Measurements (BSM) to achieve entanglement swapping and remote end-to-end entanglement establishment. Each entanglement source on the satellites has an entanglement generation rate $R_s$ and a probability of generation success denoted by $P_s$. Additionally, we denote the probability of successfully performing a BSM on repeater nodes by $P_b$. Each node in the network is endowed with a multimode quantum memory, with a coherence time $t_c$, able to store $n$ shares of EPR pairs simultaneously with a probability of failure denoted by $P_m$. We assume the probability of failure of the modes is independent. Moreover, due to the coherence time of the quantum memory, we assume that the stored states undergo additional depolarizing noise, with a noise parameter $W_m=e^{-\frac{t}{t_c}}$, where $t$ is the storage time before performing a BSM.

\section{Outage probability Utility Analysis of entanglement links and paths}
\label{sec:4}
It is clear that each link of the network comprises of two independent attributes that may cause the outage of the link: the quantum memory decoherence attribute and the EPR transmission Fidelity attribute. The outage probability of the link due to memory decoherence is simply modelled by \footnote{In many cases, the dynamics of a system undergoing dephasing can be described using a master equation, which leads to an exponential decay of coherence (often represented by the off-diagonal elements of the density matrix). This decay reflects the loss of phase relationships over time.}:
\begin{equation}
    P_{ij}^{out-m}= 1- e^{-\frac{t}{t_c}}
\end{equation}
where $t$ is the duration of the link from its establishment until an entanglement swapping using a BSM is applied.  As a matter of fact, the utility attribute of an entanglement link $x={ij}$ is given by:
\begin{align}
    U_x^{m}&= 1-P_x^{out-m}\nonumber\\
    &=e^{-\frac{t}{t_c}}
\end{align}

For the entanglement transmission fidelity attribute, different effects enters into the game. Namely, we assume that the main depolarization of the signal is due to:
\begin{enumerate}
    \item the transmittivity of the link  which scales polynomially with its distance $d_x$
    \item the pointing errors of the signal beam
    \item the turbulence associated with the atmospheric effects when considering the downlink from the repeater satellite to the ground station.
\end{enumerate}

Observing the noise model affecting the distributed Werner state, we can define a quantity, called total SNR, denoted by $SNR_T$, in terms of the noise parameter $W$ which comprises both quantum and classical effects. The total SNR is given by \footnote{The motivation behind introducing this formula linking  $SNR_T$ and the noise parameter $W$ is reflected from the idea that in a depolarizing channel:
\begin{itemize}
    \item The signal portion is preserved with probability $W$
    \item  The noise portion is introduced with probability $1-W$
\end{itemize}
Thus, the ratio between the signal and noise probabilities directly reflects how much signal is transmitted relative to how much noise is introduced, which is the essence of the SNR concept.} 
\color{black}
\begin{align}
    SNR_T&= \frac{W}{1-W}   
\end{align}
\color{black}
From the previous equations, we can easily deduce an expression of the fidelity of the distributed entangled state in terms of the corresponding SNR of the link, and it is given by:
\begin{equation}
    F=\frac{1+4SNR_T}{4+4SNR_T}
\end{equation}
We consider a link $x$ to be out of service whenever the fidelity of the transmitted EPR pair $F$ is below a given fidelity threshold $\Gamma$. This condition
translates for $SNR_T$ as:
\begin{equation}
    SNR_T\leq\frac{4\Gamma-1}{4-4\Gamma}
\end{equation}
 \color{black}
 We distinguish two different links, the downlink and the intersatellite links which have different effects on  $SNR_T$ \footnote{\color{black}We do not assume any site diversity for atmospheric effects mitigation}:

\subsubsection{Downlink}
For the downlink, the transmitivity of the link due to its distance $d_x$, the pointing errors of the signal beam and the turbulence associated with the atmospheric effects when considering the downlink from the repeater satellite to the ground station affect the $SNR_C$ in the following way:
\begin{equation}
    SNR_T=\frac{d_x(t)^{-\gamma}\langle h_x\rangle ^2P_TY_x(t)}{N_0}
\end{equation}
where $\gamma$ is the path loss factor, $\langle h_x\rangle^2$ is the covariance of the pointing error, $P_T$ is the transmission power, $Y_x$ is the factor related to atmospheric effects and $N_0$ the background noise power.

 $\langle h_x\rangle^2$ is modeled as a random variable, following a chi-square distribution  $\chi_n^2$  with $n$ degrees of freedom and mean divergence angle $\Omega$ as \cite{bouchet2010free,slim}:
\begin{equation}
    f_{\langle h_x\rangle^2}(z)=\frac{z^{\frac{n}{2}-1}e^{-\frac{z}{\Omega}}}{\Omega^{\frac{n}{2}}\Gamma\Big(\frac{n}{2}\Big)}
\end{equation}
The atmospheric factor $Y_x(t)$ is modeled using a gamma-gamma distribution as \cite{bouchet2010free}:
\begin{equation}
    f_{Y_x(t)}(y)=\frac{2(\alpha\beta)^{\frac{\alpha+\beta}{2}}}{\Gamma(\alpha)\Gamma(\beta)}y^{\frac{\alpha+\beta}{2}}K_{\alpha-\beta}(2\sqrt{\alpha\beta y})
\end{equation}
where $K_{\alpha-\beta}$ is the modified Bessel function of the second kind, with $\alpha$ and $\beta$ are parameters reflecting the strength of the turbulence depending on the Rytov variance. 
As a result, the outage probability of the downlink due to the Fidelity threshold translates as the cumulative probability given by:
\begin{equation}
    P_x^{out-fid-downlink}=P(\langle h_x\rangle^2Y_x(t)<\eta)
\end{equation}
with 
\begin{equation}
\eta=\frac{\beta N_0d_x(t)^\gamma}{P_T}
\end{equation}
and 
\color{black}
\begin{equation}
    \beta=\frac{4\Gamma-1}{4-4\Gamma}
\end{equation}
\color{black}
Assuming independence of the two PDF's $P_x^{out-fid-downlink}$ is given by:
\begin{equation}
    P_x^{out-fid-downlink}=\int_0^\infty\int_0^{\frac{\eta}{y}} f_{<h_x>^2}(h) f_{Y_x(t)}(y)dhdy
\end{equation}
The integral is complex for generic $n, \alpha$ and $\beta$, which necessitates numerical methods. 
\subsubsection{Intersatellite links:}
For the intersatellite links the turbulence due to atmospheric effects is abscent therefore the SNR formulation is simpler as we only drop the factor $Y_x$. Therefore, the outage probability of the intersatellite links, when $n=2$, is given by:
\begin{align}
  P_x^{out-fid-ISL}&=\int_0^\eta \frac{1}{\Omega}e^{-\frac{x}{\Omega}}\nonumber\\
  &= 1- e^{-\frac{\eta}{\Omega}}
\end{align}
As a result, the utility function for the ISL links is given by:
\begin{align}
    U_x^{fid-ISL}&=1-P_x^{out-fid-ISL}\nonumber\\
    &=e^{-\frac{\eta}{\Omega}}
\end{align}

Combining the two utility attributes, quantum memory and fidelity utilities, the total utility of an entanglement link $x$ is given by:
\begin{equation}
\label{weights}
    U_x=(U_x^{fid-l})^{\omega_f}\cdot (U_x^m)^{\omega_m} 
\end{equation}
with $\omega_f$ and $\omega_m$ are the weights corresponding to each utility attribute.  This link utility function, constituting the weights of the space-time network models of Fig.~\ref{illustrating figure}, is time dependent through the time dependence of the time evolution of the quantum memory attribute and the time dependence of the fidelity attribute through the time evolution of the length of the link.

For an entanglement path $EP$, the path utility $U_{EP}$ is given by:
\begin{align}
   U_{EP}&=\Pi_{x\in EP}U_x \nonumber\\
   &= \Pi_{x\in EP}(U_x^{fid-l})^{\omega_f}\cdot (U_x^m)^{\omega_m} 
\end{align}
Indeed, the entanglement path attribute depends on time, which make the choice of the contributing links to it crucial in such a way they should collectively optimize the path utility. 
\section{Space-Time Entanglement Distribution Algorithm}
\label{sec:5}
\subsection{Utility based Optimal Path Finding}
Suppose the that and end-to-end entanglement path is to be established between a source node $s$ and a destination node $d$. For instance, suppose that node $s$ during transmission is within snapshot time $t_s$ and the node $d$ is within snapshot time $t_d$ such that $t_d-t_s<t_c$. As of such, the entanglement path between the two nodes $s$ and $d$ is to be found in the virtual space-time graph representation of the network topology between the nodes $(t_s,s)$ and $(t_d,d)$ within a time horizon equals to the quantum memory coherence time $t_c$. The optimal path optimizes the utility of the path between $(t_s,s)$ and $(t_d,d)$ while taking into consideration the time dependence of the utility functions. The algorithm finding the optimal path in this time window, is given by Algorithm.~\ref{algo:optimal_path_finding}. 
\begin{algorithm}[H]
\caption{Utility based optimal Path Finding}
\label{algo:optimal_path_finding}
\begin{algorithmic}[1]
\State \textbf{Variables:}
\State \(x_{(t,i),(t',j)}\): Binary variable indicating if the edge between nodes \((t,i)\) and \((t',j)\) is part of the path.

\State \textbf{Objective:}
\begin{equation*}
\text{maximize} \quad U = \Pi_{x\in E}(U_x^{fid-l})^{\omega_f}\cdot (U_x^m)^{\omega_m} \cdot x_{(t,i),(t',j)}
\end{equation*}

\State \textbf{Constraints:}

\State \textbf{Path Continuity:}
\begin{align*}
&\sum_{(t',j) \in V} x_{(t_s, s),(t',j)} = 1 \\
&\sum_{(t,i) \in V} x_{(t,i),(t_d, d)} = 1 \quad  \\
&\sum_{(t',j) \in V} x_{(t,i),(t',j)} - \sum_{(t',j) \in V} x_{(t',j),(t,i)} = 0 \quad \\
& \forall (t,i) \neq (t_s, s), (t_d, d)
\end{align*}

\State \textbf{Binary Constraints:}
\begin{equation*}
x_{(t,i),(t',j)} \in \{0, 1\} \quad \forall (t,i),(t',j) \in E
\end{equation*}

\State \textbf{Link Fidelity Constraints:}
\begin{equation*}
F_{(t,i),(t',j)} \geq \Gamma_{min} \quad \forall (t,i),(t',j) \in E
\end{equation*}
\end{algorithmic}
\end{algorithm}

In this description, the goal is to establish an optimal path between the source node $s$ at time $t_s$ and the destination node $d$ at time $t_d$, where $t_d - t_s < t_c$. The entanglement path must be found in the virtual space-time graph representation of the network topology within a time horizon equal to the quantum memory coherence time $t_c$. The algorithm aims to find the optimal path by maximizing the utility function $U$, which is a product of the utilities of the edges in the path, taking into account the time changing distances between different nodes as well as different errors coming from mis-pointing of beams or turbulence effects. The constraints ensure path continuity, binary nature of the path variables, and a minimum link fidelity requirement.

\subsection{Optimal Entanglement Path}
When trying to find an optimal utility path between source $s$ and destination $d$, we know the initial transmission time, we know the time horizon of the path, but we do not know the snapshot time where the path is going to end. In other words, we perfectly know $(t_s,s)$, and should look for the optimal $t_d$ in $(t_d,d)$ satisfying the time horizon constraint $t_d-t_s<t_c$. Therefore, an optimal path finding algorithm as of Algorithm.~\ref{algo:optimal_path_finding} cannot be applied directly. An algorithm for optimal entanglement path finding is given in Algorithm.~\ref{algo:optimal entanglement path}.

\begin{algorithm}[H]
\caption{Optimal entanglement path}
\label{algo:optimal entanglement path}
\begin{algorithmic}[1]
\State  \textbf{Input:} Source satellite s, destination satellite d, transmission time \(t\), memory cohrent time \(t_c\)
\State \textbf{Output:} End-to-End entanglement between s and d
\State Initialize the orbit status information of satellite nodes
\State Construct the space-time evolution graph \((G,E,T)\) based on the orbital propagation rules. 
\State Set time horizon \(t_h \gets t + t_c\) 
\State optimal paths \(\gets \emptyset \)
\For{$t_l$ in $[t, t_h]$}
    \State Find the optimal path between \((t,i)\) and \((t_l, j)\) Algorithm [1]
    \State optimal paths \(\gets\) optimal path
\EndFor  
\State From optimal paths: Select the optimal path with the largest utility\\
\Return optimal entanglement path
\end{algorithmic}
\end{algorithm}

\subsection{Nested entanglement Protocol}
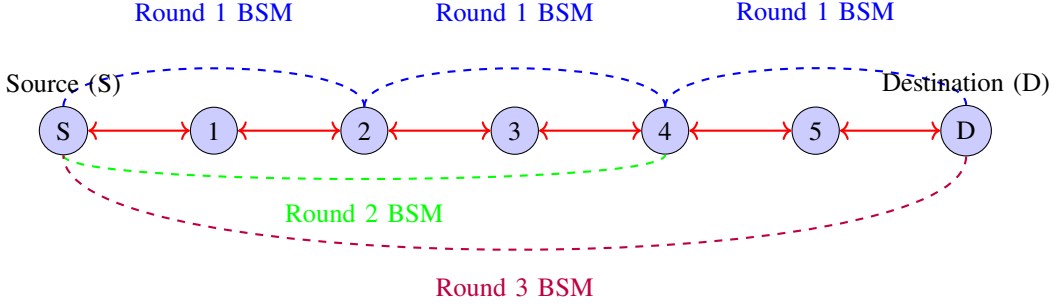
\begin{figure*}[ht]
    \centering
    \begin{tikzpicture}[scale=1.0, every node/.style={scale=1.0}]
        \node[draw, circle, fill=blue!20] (S) at (0, 0) {S};
        \node[draw, circle, fill=blue!20] (1) at (2, 0) {1};
        \node[draw, circle, fill=blue!20] (2) at (4, 0) {2};
        \node[draw, circle, fill=blue!20] (3) at (6, 0) {3};
        \node[draw, circle, fill=blue!20] (4) at (8, 0) {4};
        \node[draw, circle, fill=blue!20] (5) at (10, 0) {5};
        \node[draw, circle, fill=blue!20] (D) at (12, 0) {D};

        \draw[<->, thick, red] (S) -- (1) node[midway, above, sloped] {};
        \draw[<->, thick, red] (1) -- (2) node[midway, above, sloped] {};
        \draw[<->, thick, red] (2) -- (3) node[midway, above, sloped] {};
        \draw[<->, thick, red] (3) -- (4) node[midway, above, sloped] {};
        \draw[<->, thick, red] (4) -- (5) node[midway, above, sloped] {};
        \draw[<->, thick, red] (5) -- (D) node[midway, above, sloped] {};

        \draw[dashed, thick, blue] (S) .. controls (0, 1) and (4, 1) .. (2) node[midway, above, yshift=0.5cm] {Round 1 BSM};
        \draw[dashed, thick, blue] (2) .. controls (4, 1) and (8, 1) .. (4) node[midway, above, yshift=0.5cm] {Round 1 BSM};
        \draw[dashed, thick, blue] (4) .. controls (8, 1) and (12, 1) .. (D) node[midway, above, yshift=0.5cm] {Round 1 BSM};

        \draw[dashed, thick, green] (S) .. controls (0, -0.75) and (8, -0.75) .. (4) node[midway, below, yshift=-0.2cm] {Round 2 BSM};

        \draw[dashed, thick, purple] (D) .. controls (12, -2) and (0, -2) .. (S) node[midway, above, yshift=-0.75cm] {Round 3 BSM};

        \node[above] at (S.north) {Source (S)};
        \node[above] at (D.north) {Destination (D)};

    \end{tikzpicture}
    \caption{Illustration of the Nested Entanglement Protocol. Nodes $1, 2, 3, 4,$ and $5$ are repeater nodes performing Bell State Measurements (BSMs). In Round 1, independent BSMs are performed on nodes $1,3$ and $5$. In Round 2, BSMs are performed on node $2$. In Round 3, a BSM is performed on  node $4$. At the end of round 3 a lon haul entanglement is established between $S$ and $D$}
    \label{fig:nested_entanglement}
\end{figure*}

Let $n$ be the total number of nodes in the path $P$, and hence $n-2$ is the number of repeater nodes. The goal is to perform BSMs in a configuration that reduces the number of simultaneous operations required at any step.

The number of repeater nodes $N_{\text{repeater}} = n - 2$. At each round $i$, the number of BSMs to be performed is given by:
\begin{equation}
    N_i = \left\lceil \frac{N_{\text{remaining}}}{2} \right\rceil
\end{equation}

This process continues until the remaining repeater nodes reach zero.

The duration of completion depends on the number of rounds required to reduce $N_{\text{remaining}}$ to zero. If $T_{\text{BSM}}$ is the time taken to perform a BSM, the total duration $T_{\text{total}}$ is given by:
\begin{equation}
    T_{\text{total}} =  \sum_{i=1}^{\log_2(N_{\text{repeater}})} T_{\text{BSM}}
\end{equation}

The algorithm divides the set of repeater nodes into progressively smaller subsets, halving the number of nodes at each round. This logarithmic approach ensures that the number of rounds needed is $\log_2(N_{\text{repeater}})$, leading to efficient entanglement distribution.

The nested entanglement algorithm optimizes the entanglement swapping process, reducing the overall time complexity and ensuring robustness against failures by incorporating a retry mechanism.

\begin{algorithm}[H]
\caption{Nested Entanglement protocol}
\label{algo:nested}
\begin{algorithmic}[1]
\State \textbf{Input:} {Source $S$, Destination $D$, Entanglement Path between $S$ and $D$, number of nodes in the path $|P|$}
\State \textbf{Output:} End-to-end entanglement between $S$ and $D$
    \State $N_{\text{repeater}} \gets |P| - 2$
    \State BSM configurations $\gets []$
    \State $N_{\text{remaining}} \gets N_{\text{repeater}}$
    \While{$N_{\text{remaining}} > 0$}
        \State $N_i \gets \lceil N_{\text{remaining}} / 2 \rceil$
        \State  BSM configurations \(\gets N_i\)
        \State $N_{\text{remaining}} \gets N_{\text{remaining}} - N_i$
    \EndWhile
    \For{each $N_i$ in BSM configurations}
        \State Perform simultaneous BSMs among $N_i$ repeater nodes
    \EndFor
    \State \textbf{return} End-to-end entanglement between $s$ and $d$
\end{algorithmic}
\end{algorithm}

The fidelity of the end-to-end distributed entanglement when considering depolarizing noise in the transmission and in the quantum memories, and when considering the previousely discussed nested entanglement protocol, is given by:

\begin{widetext}
\begin{equation}
\label{fidelity}
    F = \frac{1 + 3 \Pi_{x=1}^{n-1} W_x \Pi_{i=1}^{\lceil \log_2 N_{\text{repeater}} \rceil} W_m(\sum_{l=1}^i \delta_l)^{2N_i} \cdot W_m(\sum_{l=1}^{\lceil \log_2 N_{\text{repeater}} \rceil} \delta_l)^2}{4}
\end{equation}
\end{widetext}

where $\{\delta_i\}_{i=1}^{\lceil \log_2 N_{repeater} \rceil}$ are the durations for completing the $\lceil \log_2 N_{repeater} \rceil$ rounds of the nesting protocol
(See Appendix.~\ref{appendix1} for details)\\
\\
The long haul entanglement distirbution success relies of the success of each source to generate an entangled pair, BSM to perform entanglement swapping and quantum memory to store the share of entanglement. The probability of entanglement distribution success in a chain of $n$ nodes is given by
\begin{equation}
    \left(P_s^{(n-1)}\right) \left(\left(1-P_m^{\text{X}}\right)^{2(n-1)}\right) \left(P_b^{(n-2)}\right)
\end{equation}
 where $X$ is the possible number of possible modes that can be stored in the quantum memory. We are considering that the quantum memory fails only if all the modes fail to store entanglement.
 
 The end to end entanglement throughput $r$ in this case, if we consider an entanglement generation rate and storage rate equal to $R$ is given by:
 \begin{equation}
 \label{throughput}
     r=R\cdot \left(P_s^{(n-1)}\right) \left(\left(1-P_m^R\right)^{2((n-1)}\right) \left(P_b^{(n-2)}\right)\cdot \Pi_{x=1}^{n-1} \eta^t_{x}
 \end{equation}
where $\eta_x^t$ is the transmittance of the link $x$.
\subsection{Space-time entanglement distribution algorithm}
Finding the optimal entanglement path requires a sophisticated entanglement distribution strategy. In this paper, we consider nested entanglement distribution, which necessitates a strategic approach to Bell State Measurements (BSM) for entanglement swapping. This strategy is crucial for optimizing the resources along the path, particularly by mitigating the effects of memory decoherence time, which can drastically affect the fidelity of the ultimately distributed entangled state.

When establishing an optimal entanglement path, it is common to encounter subpaths that belong to the overall path but exist entirely in different snapshots. If we wait to perform nested entanglement distribution until the entire point-to-point entanglement distribution is completed across all snapshots—from the initial to the final snapshot—even within the coherence time of the quantum memory, the fidelity of the distributed entangled pair would be reduced and is given, stemming from Eq.~\ref{fidelity}, by:

\begin{widetext}
\begin{equation}
\label{fidelity2}
    F = \frac{1 + 3 \Pi_{x=1}^{n-1} W_x \Pi_{i=1}^{\lceil \log_2 N_{\text{repeater}} \rceil} W_m((t_f-t_i)+\sum_{l=1}^i \delta_l)^{2N_i} \cdot W_m(\sum_{l=1}^{\lceil \log_2 N_{\text{repeater}} \rceil} \delta_l)W_m((t_f-t_0)+\sum_{l=1}^{\lceil \log_2 N_{\text{repeater}} \rceil} \delta_l)}{4}
\end{equation}
\end{widetext}
where $\{t_i\}_{i=0}^f$ are the begining of each snapshot involved in the distirbution path.This reduction occurs because of the extended waiting times required to perform the nesting BSMs at the end.
To address this issue, it is advantageous to implement segmented nesting along the different snapshots. By performing nested entanglement distribution incrementally across different snapshots, we can significantly reduce the waiting times and thereby minimize the decoherence effects. This approach ensures higher fidelity of the distributed entangled pairs and better utilization of the network resources. 
Algorithm.~\ref{algo:STBD} outlines the segmented nesting strategy, demonstrating its superiority over the wait-until-the-end approach in maintaining higher end-to-end entanglement fidelity. This strategy involves performing BSMs at intermediate stages along the path, thereby optimizing the overall distribution process and enhancing the robustness of the quantum network. The fidelity for the segmented nesting protocol is given by:
\begin{equation}
\label{}
    F = \frac{1 + 3 W_{SBTD}}{4}
\end{equation}
with $W_{SBTD}$ is given by:

\section{Analysis}
\label{sec:6}
\begin{figure}
    \centering
    \includegraphics[width=0.9\columnwidth]{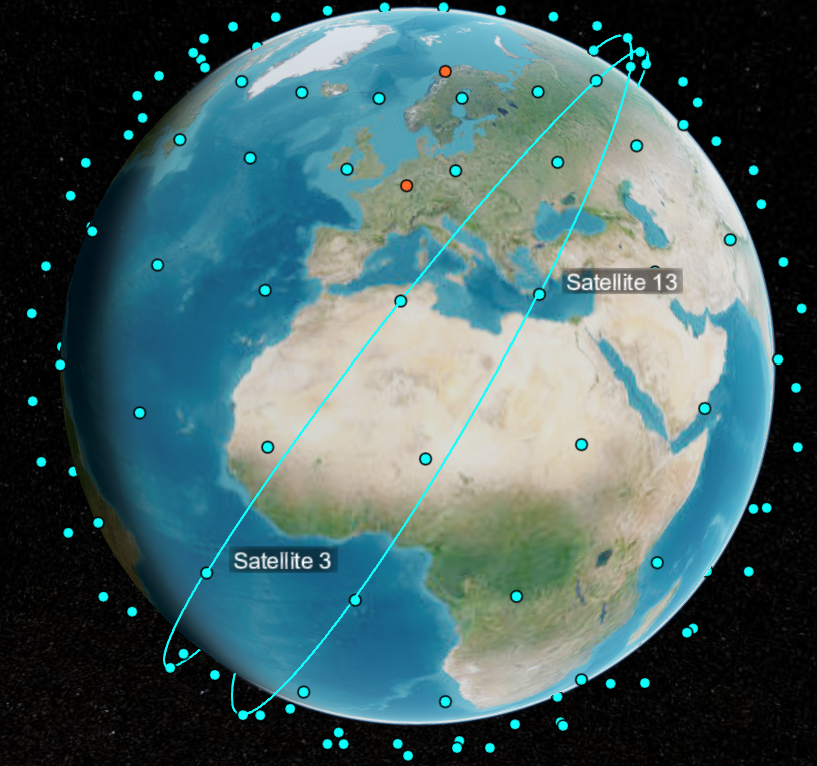}
    \caption{An illustration of the LEO constellation highlighting satellites 3 and 13 and the ground stations in Luxembourg and Norway}
    \label{constellation}
\end{figure}
In this study, we analyze the performance of entanglement distribution algorithms in a Low Earth Orbit (LEO) Walker satellite constellation network. The focus is on ground-to-ground space-based entanglement distribution between Luxembourg and Norway, as well as satellite-to-satellite entanglement between satellites 3 and 13 within the constellation as highlighted in Figure.~\ref{constellation}. The goal is to understand the efficacy of these algorithms in the context of a space-time virtual topology and to evaluate the quality and reliability of entanglement distribution under the given simulation parameters.

The LEO constellation parameters, as outlined in Table \ref{tab:leo_params}, consist of 160 satellites distributed across 20 planes, with 8 satellites per plane, with the constellation orbits are at an altitude of 550 kilometers with an inclination of 60 degrees \footnote{\color{black} For this analysis, we are considering ISL's based on the line of sight and a threshold intersatellite distance equl to 5000km including inter-plane and intra-plane links. The same is considered for the downlinks}. The Earth’s radius is considered to be 6371 kilometers. Additionally, there are two ground stations involved in the network. These parameters ensure a comprehensive coverage and frequent revisit times over the ground stations, which is crucial for sustained entanglement distribution.

\begin{table}[htbp]
    \centering
    \caption{LEO Constellation Simulation Parameters}
    \begin{tabular}{|c|c|}
        \hline
        \textbf{Parameter} & \textbf{Value} \\
        \hline
        Number of Satellites & \(160\) \\
        Number of Planes & \(20\) \\
        Satellites per Plane & \( 8\) \\
        Inclination & $60^\circ$\\
        Earth Radius & \(6371 \, \si{\kilo\meter}\) \\
        LEO Altitude & \(500 \, \si{\kilo\meter}\) \\
        Number of Ground Stations & \(2\) \\
        \hline
    \end{tabular}
    \label{tab:leo_params}
\end{table}

The entanglement distribution parameters presented in Table \ref{tab:entanglement_params} play a vital role in determining the performance of the entanglement distribution algorithms. The entanglement rate is set at 10 Ebits/s, with key probabilities such as \(P_m\) (probability of failure of a single mode of the quantum memory) and \(P_b\) (BSM success probability) both at 0.1 and 0.9 respectively, and \(P_s\) (success probability of the entanglement source) also at 0.9. These high probabilities indicate a robust mechanism for creating and maintaining entangled states across the network.

\begin{table}[htbp]
    \centering
    \caption{Entanglement Distribution Parameters}
    \begin{tabular}{|c|c|}
        \hline
        \textbf{Parameter} & \textbf{Value} \\
        \hline
        Entanglement Rate & \(10 \, \text{Ebits/s}\) \\
        \(P_m\) & \(0.1\) \\
        \(P_b\) & \(0.9\) \\
        \(P_s\) & \(0.9\) \\
        Path Loss Exponent (\(\gamma\)) & \(2.0\) \\
        Transmit Power (\(P_T\)) & \(50 \, \si{\watt}\) \\
        Noise Power (\(N_0\)) & \(1 \times 10^{-9} \, \si{\watt}\) \\
        Mean of Exponential Distribution (\(\Omega\)) & \(1.0\) \\
        Fidelity Threshold Parameter (\(\Gamma\)) & \(0.8\)\\   Chi-squared distribution parameter (\(n\)) & \(2\)
        \\
        Gamma-gamma Parameter (\(\alpha\)) & \(2.1\) \\
        Gamma-gamma Parameter (\(\beta\)) & \(2.1\) \\
        Chi-Squared Distirbution parameter (\(n\)) & \(2\) \\
        Weight of SNR (\(w_f\)) & \(0.5\) \\
        Weight of Memory (\(w_m\)) & \(0.5\) \\
        \hline
    \end{tabular}
    \label{tab:entanglement_params}
\end{table}

The path loss exponent (\(\gamma\)) is set to 2.0, which reflects the free-space path loss characteristic, while the transmit power (\(P_T\)) is 50 watts and the noise power (\(N_0\)) is \(1 \times 10^{-9} \, \si{\watt}\). The mean of the exponential distribution (\(\Omega\)) is 1.0, which models the variability in link reliability. The fidelity threshold parameter (\(\Gamma\)) is 0.8, ensuring that only high-fidelity entanglement is considered successful. The gamma-gamma parameters (\(\alpha\) and \(\beta\)) are both set to 2.1, which model the atmospheric turbulence effects, and pointing errors probability distirbution function with parameter (\(n\)) is 2, indicating the number of independent spatial paths for the entangled photons.

The weights for SNR (\(w_f\)) and memory (\(w_m\)) are both set at 0.5, signifying an equal importance given to signal quality and memory effects in the entanglement distribution process. This balanced approach ensures that the algorithms account for both the immediate quality of the entangled states and the long-term retention of entanglement across the network.

By simulating these parameters, the analysis aims to provide insights into the feasibility and performance of entanglement distribution in a realistic LEO satellite constellation while considering its time varying topology. The ground-to-ground entanglement distribution between Luxembourg and Norway will be analyzed to understand the impact of ground station locations and atmospheric conditions on the entanglement quality. Similarly, the satellite-to-satellite entanglement distribution between satellites 2 and 11 will shed light on the intra-constellation dynamics and the effectiveness of the entanglement distribution protocols in space. The results from these simulations will be crucial for the future development of quantum communication networks, providing a pathway towards a global quantum internet.

\subsection{Entanglement drop rate}
\begin{algorithm}
\caption{Space-time entanglement distribution algorithm}
\label{algo:STBD}
\begin{algorithmic}[1]
\State \textbf{Input:} Source satellite \(s\), destination satellite \(d\), transmission time \(t\), memory coherence time \(t_c\)
\State \textbf{Output:} end-to-end entanglement between $s$ and $d$
\State P \(\gets\) optimal entanglement path \(((s, d, t,t_c))\)
\State snapshots \(\gets\) \text{segment path by snapshots}(P)
\State subpaths \(\gets \emptyset\) 
\State transition points \(\gets \emptyset\) 
\For{each snapshot $S_i$ in $snapshots$}
    \State subpaths \(\gets\) \text{identify subpath within snapshot}($S_i$, P)
    \EndFor
\For {$i$ in range(subpaths)}
     \State transition points \(\gets\) subpaths$[i][-1][-1]$
\EndFor
     
\For {$i$ in range $(1, len(snapshots)+1)$}
    \State \textbf{if} $i=1$ \textbf{do}
          \State \quad nested entanglement protocol (subpaths[0])
    \State \textbf{else} 
    \State subpath [i-1] \(\gets\)  subpath [i-1] $\cup$ $(s,$transition point $[i-1]$)
    \State nested entanglement protocol (subpath $[i-1]$)
\EndFor
\State \textbf{return} end-to-end entanglement between $s$ and $d$
\end{algorithmic}
\end{algorithm}
\begin{figure*}[t]
\begin{minipage}[c] {0.49\textwidth}
      \centering
    \includegraphics[width=1\columnwidth]{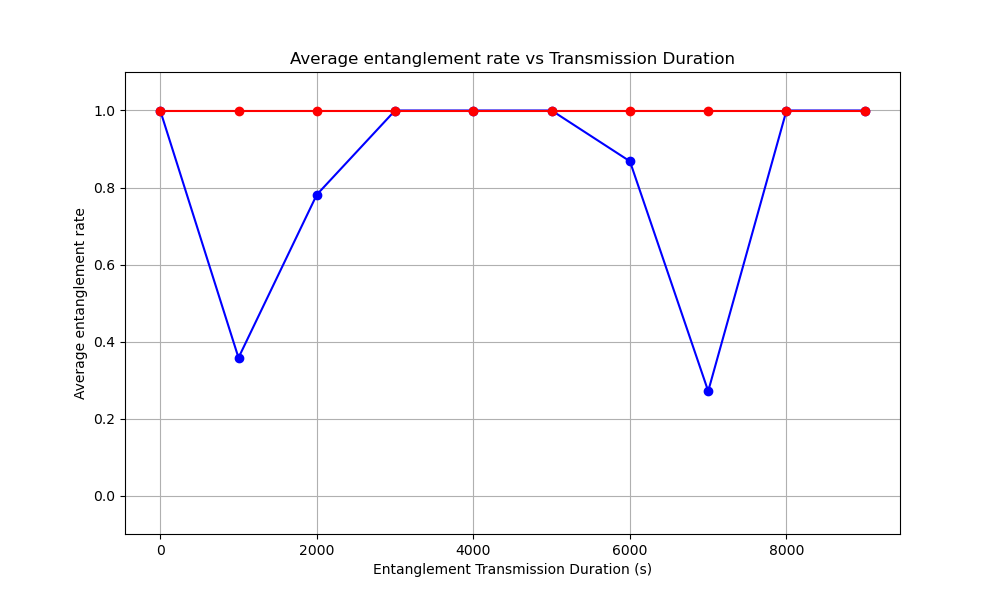}
    \subcaption{Entanglement drop rate for entanglement distribution between Luxembourg and Norway}
    \label{fig:drop rate}  
\end{minipage}
 \hspace{0.02\textwidth}
\begin{minipage}[c] {0.49\textwidth}
    \centering
    \includegraphics[width=1\columnwidth]{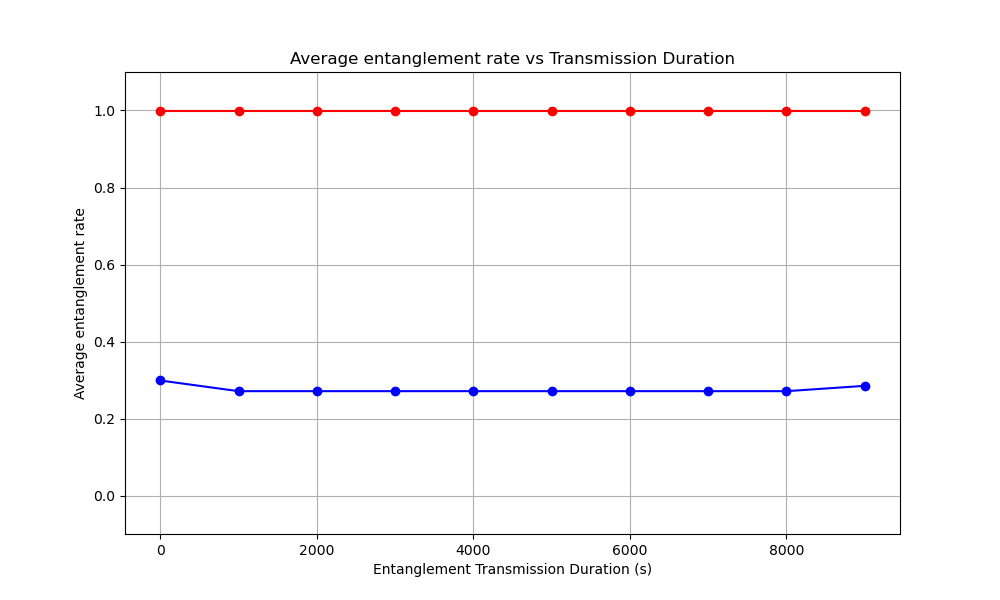}
    \subcaption{Entanglement drop rate for entanglement distribution between two satellites in the LEO constellation}
    \label{fig:drop rate satellite}
\end{minipage}
    \caption{Entanglement drop rate in space based ground-to-ground  and satellite-to-satellite entanglement distribution}
    \label{Fig:drop rate}
    \hrulefill
\end{figure*}
Entanglement drop rate is a critical figure of merit in entanglement-based quantum networks, as it quantifies the frequency at which entangled states are lost during the distribution process. A high entanglement drop rate directly impacts the reliability and efficiency of quantum communication, as it leads to frequent disruptions and necessitates additional resources to regenerate the lost entanglement. In the context of quantum LEO networks, traditional distribution algorithms often suffer from high entanglement drop rates due to several factors. The dynamic nature of satellite orbits introduces frequent changes in the communication links, leading to interruptions and misalignments. Additionally, the long distances involved and the presence of atmospheric disturbances further exacerbate the drop rate. These challenges highlight the need for optimized entanglement distribution algorithms that can mitigate the high drop rates and ensure a more stable and reliable quantum communication network.

The simulation of the entanglement drop rate both for satellite-to-satellite and ground-to-ground space based entanglement  distribution is presented in Fig.~\ref{Fig:drop rate}. The simulation is carried over different transmission times. The plots show that the STBD performs significantly better than traditional dynamic distribution \cite{dynamic}. For the ground-to-ground entanglement distribution, one can remark that entanglement drop rate increases for the STBD for different transmission duration. This is indeed due to the fact the LEO constellation simulated is not optimized to guarantee optimal coverage of the two ground stations in Luxembourg and Norway. Additional optimization of the constellation would lead naturally to better performance. In the contrary, the performance of the STBD in satellite-to-satellite entanglement distribution is significant.

\subsection{End-to-end entanglement throughput}
\begin{figure*}[t]
\begin{minipage}[c] {0.49\textwidth}
      \centering
    \includegraphics[width=1\columnwidth]{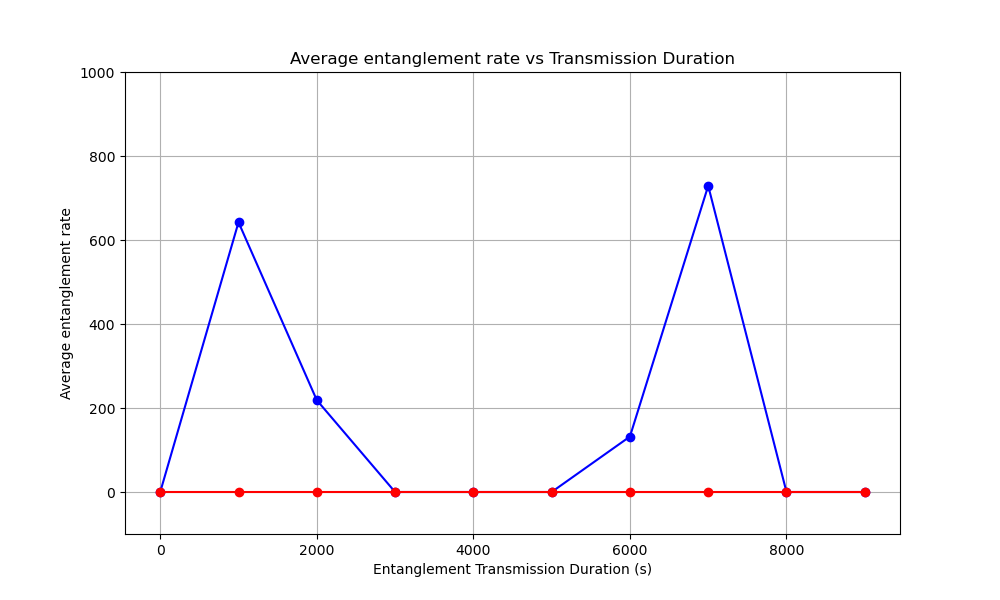}
    \subcaption{End-to-end entanglement throughput  for entanglement distribution between Luxembourg and Norway}
    \label{fig:throughput}  
\end{minipage}
 \hspace{0.02\textwidth}
\begin{minipage}[c] {0.49\textwidth}
    \centering
    \includegraphics[width=1\columnwidth]{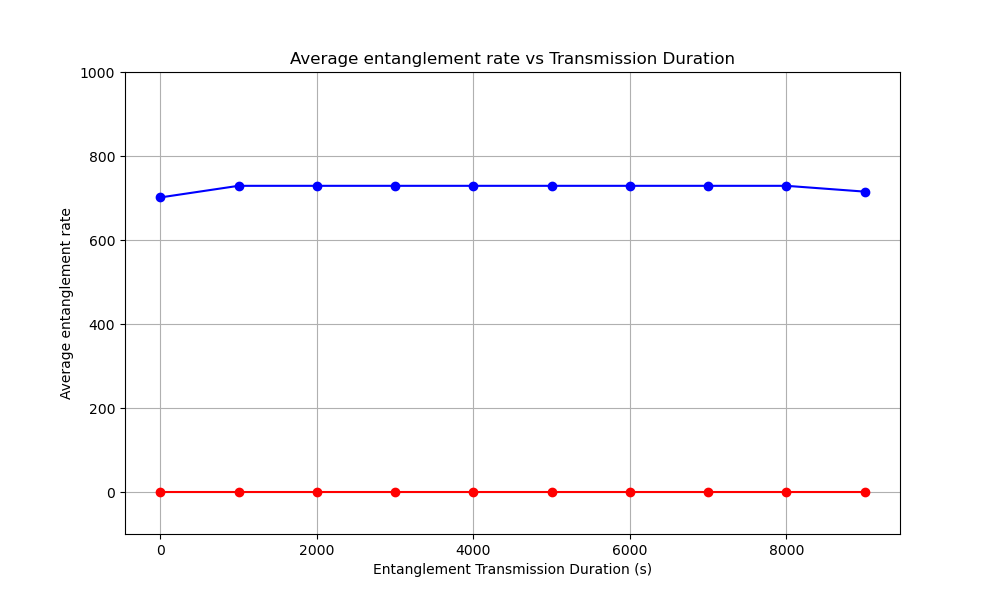}
    \subcaption{End-to-end entanglement throughput  for entanglement distribution between two satellites in the LEO constellation}
    \label{fig:throughput satellite}
\end{minipage}
    \caption{End-to-end entanglement throughput in space based ground-to-ground  and satellite-to-satellite entanglement distribution}
    \label{Fig:throughput rate}
    \hrulefill
\end{figure*}
End-to-end entanglement throughput is a crucial figure of merit in entanglement-based quantum networks, as it measures the rate at which entangled states are successfully distributed between end nodes over the network. High end-to-end entanglement throughput is essential for the practical implementation of quantum communication protocols, as it ensures that a sufficient number of entangled pairs are available for tasks such as quantum cryptography, quantum computing and quantum sensing. In the context of quantum LEO networks, achieving high throughput is challenging due to the dynamic nature of satellite movements, long distances, and atmospheric disturbances. Traditional distribution algorithms, even the dynamic ones, often struggle to maintain high throughput under these conditions, resulting in inefficient use of network resources and lower overall performance. Therefore, the development of advanced distribution algorithms that can enhance end-to-end entanglement throughput is vital for the realization of robust and efficient quantum communication networks.

The simulation results for end-to-end entanglement throughput given in Eq.~\ref{throughput} in both satellite-to-satellite and ground-to-ground space-based entanglement distribution are presented in Fig.~\ref{Fig:throughput rate},  with  the transmitance of the link $x$ is given by:
\begin{equation}
    \eta_x^t (downlink) = d_x(t)^{-\gamma} \cdot \langle h_x \rangle^2 \cdot Y_x(t)
\end{equation}
for the downlink or by 
\begin{equation}
    \eta_x^t (ISL) = d_x(t)^{-\gamma} \cdot \langle h_x \rangle^2 
\end{equation}
for the ISL links.

The simulations were conducted over various transmission times. The results indicate that the STBD algorithm significantly outperforms traditional dynamic routing. For ground-to-ground entanglement distribution, it is observed that the entanglement throughput decreases for the STBD over different transmission durations. This decline is attributed to the lack of optimization in the LEO constellation for ensuring optimal coverage of the two ground stations in Luxembourg and Norway. Further optimization of the constellation would naturally lead to improved performance. Conversely, the performance of the STBD in satellite-to-satellite entanglement distribution is notably superior.

\subsection{Fidelity}
\begin{figure*}[t]
\begin{minipage}[c] {0.49\textwidth}
      \centering
    \includegraphics[width=1\columnwidth]{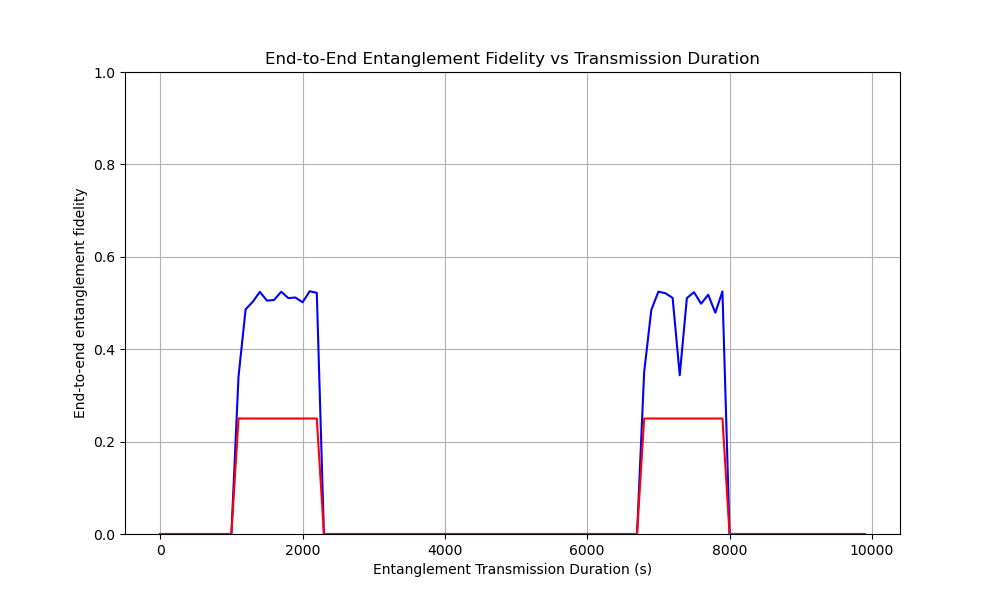}
    \subcaption{Fidelity of end-to-end entanglement between Luxembourg and Norway}
    \label{fig:fidelity}  
\end{minipage}
 \hspace{0.02\textwidth}
\begin{minipage}[c] {0.49\textwidth}
    \centering
    \includegraphics[width=1\columnwidth]{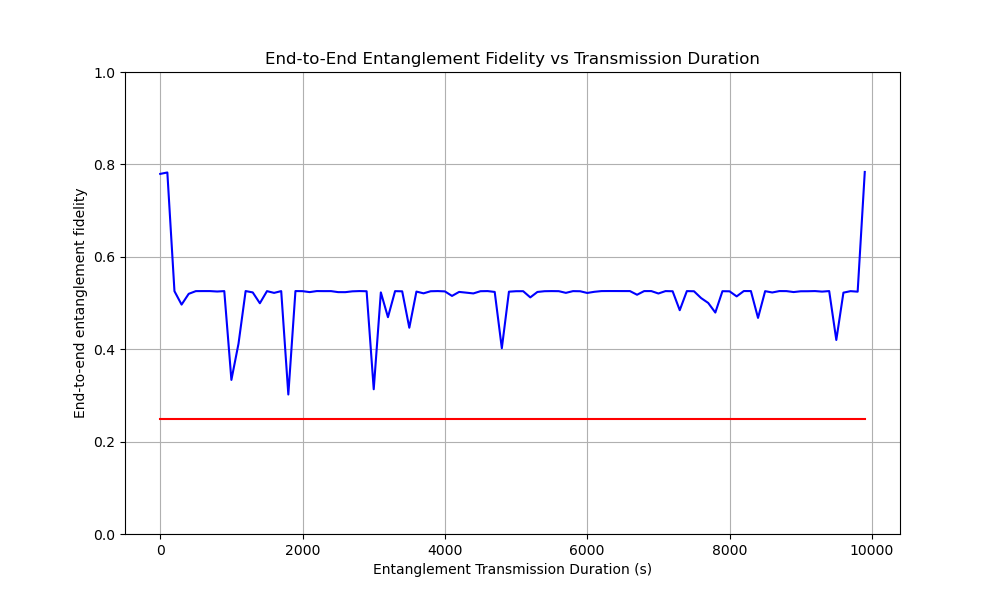}
    \subcaption{Fidelity of end-to-end entanglement between two satellites in the LEO constellation}
    \label{fig:fidelity satellite}
\end{minipage}
    \caption{Fidelity of end-to-end entanglement in space based ground-to-ground  and satellite-to-satellite entanglement distribution}
    \label{Fig:fidelity}
    \hrulefill
\end{figure*}

End-to-end entanglement fidelity is a crucial figure of merit in entanglement-based quantum networks, as it measures the overall quality of the entangled state distributed between the end nodes across the entire network. Unlike point-to-point entanglement fidelity, which only considers individual links, end-to-end fidelity takes into account the cumulative effect of all the links and nodes in the distribution path. High end-to-end fidelity is essential for various quantum applications, such as quantum key distribution, quantum teleportation, and distributed quantum computing, which require reliable and high-quality entangled states.

In quantum LEO networks, achieving good end-to-end fidelity is particularly challenging due to the dynamic nature of satellite orbits, long distances, and atmospheric disturbances. 

Implementing efficient strategies is necessary for distributing entanglement throughout different snapshots in the transmission duration, as the optimal paths can change rapidly due to the movement of the satellites. By focusing on end-to-end fidelity, it is possible to maintain high-quality entangled states over the entire network, ensuring the reliability and efficiency needed for various quantum communication applications. 

The plots of fidelity for both ground-to-ground and satellite-to-satellite scenarios are presented in Fig.~\ref{Fig:fidelity}. The red curve represents the fidelity when nested entanglement distribution is performed only after the entire entanglement path is established among all snapshots following Eq.~\ref{fidelity2}. In contrast, the blue curve depicts the fidelity for segmented nested entanglement distribution following Eq.~\ref{fidelity-segmented}. The segmented nesting approach demonstrates significantly better performance than the wait-until-the-end nesting approach. This improvement is due to the reduction of noise caused by time evolution during storage in quantum memories. By segmenting the entanglement distribution process, the impact of decoherence and other noise factors is minimized, resulting in higher overall fidelity for the distributed entangled states.

While the segmented nesting approach significantly increases end-to-end entanglement fidelity, some distribution paths still experience low fidelity levels. For applications requiring high fidelity thresholds, these low-fidelity entangled states would be discarded. Additionally, the fidelity of entanglement distributed between satellite nodes is notably higher on average than that between ground stations. This difference is mainly due to atmospheric turbulence affecting the downlink, which is absent in intersatellite links. Furthermore, fidelity drops to zero for ground-to-ground transmission when no entanglement paths are available within the corresponding transmission time.

These fidelities are also highly influenced by the quantum memory coherence time. In the following sections, we investigate the impact of quantum memory coherence time on the quality of end-to-end entanglement fidelity using STBD and segmented nested entanglement distribution. We also examine its effects on entanglement throughput and entanglement drop rate.
\begin{figure*}
    \begin{equation}
    \label{fidelity-segmented}
        \Pi_{x=1}^{n-1} W_x \Pi_{k=1}^K\Pi_{i=1}^{{\lceil \log_2 N_{\text{repeater}}^k\rceil}} W_m(\sum_{l=1}^i\delta^k_l)^{2N_i^k}W_m((t_k-t_{k-1})+\sum_{l=1}^{\lceil \log_2 N_{\text{repeater}}^k\rceil}\delta^k_l) \cdot W_m(t_f-t_0+\sum_{k=1}^{K}\sum_{l=1}^{\lceil \log_2 N_{\text{repeater}}^k\rceil} \delta_l) W_m(\delta_{\lceil \log_2 N_{\text{repeater}} \rceil})
    \end{equation}
\end{figure*}

\subsection{The effect of quantum memories coherence time on space-based entanglement distirbution}
\begin{figure*}[t]
\begin{minipage}[c] {0.49\textwidth}
      \centering
    \includegraphics[width=1\columnwidth]{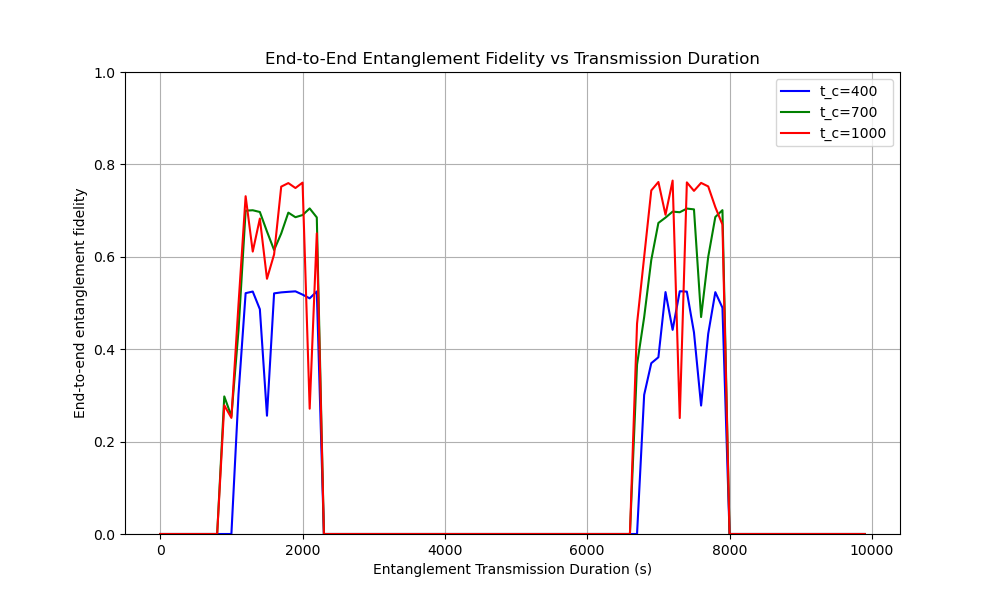}
    \subcaption{The behaviour of the fidelity of end-to-end entanglement between Luxembourg and Norway with respect to different coherence times}
    \label{fig:fidelity-memory-groun}  
\end{minipage}
 \hspace{0.02\textwidth}
\begin{minipage}[c] {0.49\textwidth}
    \centering
    \includegraphics[width=1\columnwidth]{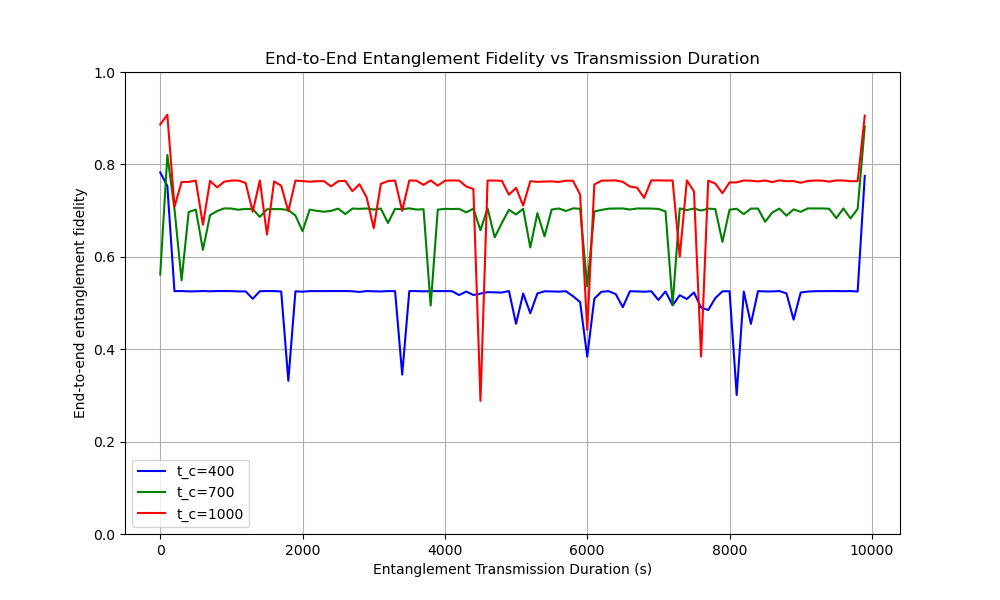}
    \subcaption{The behaviour of the fidelity of end-to-end entanglement between two satellites in the LEO constellation with respect to different coherence times}
    \label{fig:fidelity-memrory- satellite}
\end{minipage}
    \caption{The behaviour of the fidelity of end-to-end entanglement in space based ground-to-ground  and satellite-to-satellite entanglement distribution, using segmented nesting, with respect to quantum memories coherence times}
    \label{Fig:fidelity-memory}
    \hrulefill
\end{figure*}

Quantum memories are a crucial component in quantum communication systems, acting as storage devices for quantum information. However, the manufacturing of quantum memories faces several technological challenges. One of the primary limitations is the coherence time, which is the duration for which a quantum state can be preserved without significant degradation. Current quantum memories exhibit coherence times that are often insufficient for many practical applications, particularly those involving long-distance entanglement distribution. Short coherence times lead to increased decoherence and noise, resulting in lower fidelity of the stored and retrieved quantum states.

Advancements in materials science and quantum engineering are necessary to extend the coherence times of quantum memories. Techniques such as dynamical decoupling, error correction, and the use of ultra-pure materials are being explored to mitigate decoherence effects. Increasing the coherence time is of paramount importance for quantum communications because it directly impacts the performance and reliability of quantum networks. Longer coherence times allow for more extended and stable storage of entangled states, enabling efficient entanglement swapping and purification processes essential for long-distance quantum communication. As research progresses, overcoming the technological limits of quantum memories will be critical to realizing robust and scalable quantum communication networks. Recent developement on quantum memories have reported a coherence time of an $^{171}$Yb$^+$ ion-qubit  up to $5500s$ \cite{wang2021single}.
In Fig. ~\ref{Fig:fidelity-memory} is reported the behavour of the fidelity of the distributed end-to-end entanglement using segmented nesting within the snapshots. It is clear from both plots, that the increase of the coherence time of quantum memories increases significantly the performance of the space-time based entanglement distribution in the LEO network, with nearly above 50\% in mostly all the transmission times between the previous behaviour of $300s$ of coherence time reported in Fig.~\ref{Fig:fidelity}, and $1000s$ of coherence time reported in Fig.~\ref{Fig:fidelity-memory}. s well, we can notice from Fig.~\ref{fig:fidelity-memory-groun} that there are new paths appearing that did not exist a priori in the case of coherence time of $300s$ due to the increase of the time horizon of Algorithm.~\ref{algo:optimal entanglement path}

\section{Conclusions}
In this paper, we have introduced a novel framework for entanglement distribution in LEO satellite constellations, leveraging space-time graphs to effectively manage the dynamic topology of such networks. Our approach represents a significant advancement in the field of quantum communications, addressing the inherent challenges posed by the time-varying nature of satellite orbits.

The use of space-time graphs allows us to model the temporal evolution of the network, enabling a more reliable and efficient method for entanglement distribution. By optimizing path utility that incorporates factors such as pointing errors, non-ideal link transmittance for intersatellite connections, and atmospheric effects for downlinks, our strategy significantly reduces entanglement drop rates and enhances end-to-end entanglement throughput.

A critical aspect of our approach is its dependency on the coherence times of quantum memories. Segmented nested entanglement distribution has been shown to outperform the -wait till the end- approach of nesting. Additionally, longer coherence times facilitate improved performance, thereby making the proposed method even more effective. This highlights the importance of developing advanced quantum memories with extended coherence times to maximize the benefits of our scheme.

The implications of our work are profound for the future of the space-based quantum internet. By transforming the challenges of a time-varying topology into opportunities, we pave the way for the optimal use of LEO satellite constellations in distributing entanglement reliably between distant ground or space users. This reliability and efficiency are crucial for realizing various applications such as distributed quantum computing, quantum cryptography, and quantum sensing on a global scale. Our research provides a foundational step towards understanding and harnessing the potential of satellite-based quantum networks, bringing us closer to the practical implementation of a robust and secure quantum internet.

\section*{Acknowledgement}
This work was supported by the project Lux4QCI (GA
101091508) funded by the Digital Europe Program, and the
project LUQCIA Funded by the European Union – Next Generation EU, with the collaboration of the Department of Media, Connectivity and Digital Policy of the Luxembourgish Government in the framework of the RRF program. S.K thanks Mert Bayraktar and Muhammad Ahsan for discussion.

\small
\bibliographystyle{IEEEtran}
\bibliography{ref}

\vspace{50pt}
\appendices

\section{Details of Eq.~\ref{fidelity}}
\label{appendix1}
First of all we show that the concatenation of two depolarizing channels $\mathcal{N}_{W_1}$ and $\mathcal{N}_{W_2}$ with parameters $W_1$ and $W_2$ respectively, is identical to a depolarizing channel $\mathcal{N}_{W_3}$ with depolarizing parameter $W_3 = W_1 \cdot W_2$. For instance:
\begin{equation}
    \mathcal{N}_{W_1}\big(\ketbra{\Phi^+}{\Phi^+}\big)=W_1\ketbra{\Phi^+}{\Phi^+}+(1-W_1)\frac{\mathrm{I}}{4} 
\end{equation}
where it is understood that the channel is acting on one of the qubits of the EPR pair $\ketbra{\Phi^+}{\Phi^+}$, be it the first qubit.
By letting the second channel act we have:
\begin{align}
    \mathcal{N}_{W_2}\circ\mathcal{N}_{W_1}\big(\ketbra{\Phi^+}{\Phi^+}\big)&=\mathcal{N}_{W_2}\Big(W_1\ketbra{\Phi^+}{\Phi^+}+(1-W_1)\frac{\mathrm{I}}{4} \Big)\nonumber\\
    &=W_2W_1\ketbra{\Phi^+}{\Phi^+}+W_1(1-W_2)\frac{\mathrm{I}}{4}+\nonumber\\
    & +W_2(1-W_1)\frac{\mathrm{I}}{4}+(1-W_2)(1-W_1)\frac{\mathrm{I}}{4}\nonumber\\
    &= W_1W_2 \ketbra{\Phi^+}{\Phi^+} + (1-W_1W_2)\frac{\mathrm{I}}{4}\nonumber
\end{align}
therefore $\mathcal{N}_{W_2}\circ\mathcal{N}_{W_1}=\mathcal{N}_{W_3}$ with $W_3=W_1W_2$.
We note that the same result still holds if the two channels are not concatenated but act on different qubits. This is formally given by: 
\begin{align}
    \mathcal{N}_{W_2}\otimes\mathcal{N}_{W_1}&=\mathcal{N}_{W_3}\otimes \mathrm{I} \nonumber\\
    &=\mathrm{I}\otimes \mathcal{N}_{W_3}
\end{align}
with $W_3=W_1W_2$.
As a result, if we have $n$ depolarizing channels with parameters $\{W_i\}_{i=1}^n$ acting on a given state either on different parts or on the same part sequentially, the channel is effectively equivalent to:
\begin{equation}
    W(n) \rho + (1-W(n))\frac{\mathrm{I}}{2^k}
\end{equation}
where $k$ is the number of qubits in the state $\rho$, and $W(n)=\prod_{i=1}^n W_i$. 
Stemming from these observations and results, we can see that if we have $n-1$ EPR links $\ketbra{\Phi^+}{\Phi^+}^{\otimes (n-1)}$, with $l$ depolarizing channels acting on them either jointly or sequentially or in a combination of both, it is effectively equivalent to
\begin{equation}
    \prod_{i=1}^l W_i \ketbra{\Phi^+}{\Phi^+}^{\otimes n}+(1-\prod_{i=1}^l W_i)\frac{\mathrm{I}}{2^{n-1}}
\end{equation}
After performing successful BSM's on the EPR links, the end state fidelity is given by:
\begin{equation}
    F=\frac{1+3\prod_{i=1}^l W_i}{4}
\end{equation}
For an entanglement distribution scenario in a repeater chain of $n$ nodes, there are $n-1$ depolarizing channels acting on the different qubits during point-to-point distribution of the EPR. For each repeater node, there are two depolarizing channels acting on the two local quantum memories and there is one depolarizing noise acting on the end nodes, therefore there are $2(n-2)+2=2(n-1)$ in total. 
Consider a nested entanglement protocol on a chain where the number of repeater nodes is $N_{repeater}$. According to Algorithm \ref{algo:nested}, for each round $i$ there are $N_i$ repeater stations going to perform a BSM simultaneously on the stored entanglement in the quantum memories. If we assume that the time required to complete these BSM's successfully is $\delta_i$, therefore the quantum memories at round $l$ of nesting have been undergoing time $\sum_{i=1}^l \delta_i$, which yields a depolarization factor $W_m(\sum_{i=1}^l \delta_i)^{2N_i}$ where the power two is due to the fact that there are two memories in each repeater station. Considering all the possible rounds given by $\lceil \log_2(N_{repeater})\rceil$, we have a factor of $\prod_{i=1}^{\lceil \log_2(N_{repeater})\rceil} W_m(\sum_{i=1}^l \delta_i)^{2N_i}$. Additionally, the end nodes have undergone both evolution time equal to $t=\sum_{i=1}^{\lceil \log_2(N_{repeater})\rceil} t_i$, therefore another factor is given by $W_m(\sum_{i=1}^{\lceil \log_2(N_{repeater})\rceil} \delta_i)^2$. Summing up, the global effective factor for depolarization due to quantum memory and point-to-point transmission decoherence is:
\begin{equation}
    \prod_{x=1}^{n-1} W_x \prod_{i=1}^{\lceil \log_2 (N_{repeater})\rceil} W_m(\sum_{i=1}^l \delta_i)^{2N_i} \cdot W_m(\sum_{i=1}^{\lceil \log_2(N_{repeater})\rceil} \delta_i)^2 
\end{equation}

\end{document}